\documentclass{INTERSPEECH2023}
\usepackage{float}
\usepackage[section]{placeins}
\usepackage{subcaption} 
\usepackage{amsmath,amssymb,stmaryrd}
\usepackage{multirow} 
\usepackage{array}
\usepackage{tabularx}
\newcolumntype{x}[1]{>{\centering \arraybackslash \hspace {0pt}}p {#1}}

\interspeechcameraready

\title{Visualizing data augmentation in deep speaker recognition}
\name{Pengqi Li$^{1,3}$, Lantian Li$^2$, Askar Hamdulla$^1$, Dong Wang$^{3*}$
\thanks{
This work was supported by the National Natural Science Foundation of China (NSFC) under Grants No.62171250.}}
\address{
  $^1$School of Information Science and Engineering, Xinjiang University, China \\
  $^2$School of Artificial Intelligence, Beijing University of Posts and Telecommunications, China \\
  $^3$Center for Speech and Language Technologies, Tsinghua University, China}
\email{$^{*}$Corresponding author:~wangdong99@mails.tsinghua.edu.cn}
\begin{document}

\maketitle

\begin{abstract}

Visualization is of great value in understanding the internal mechanisms of neural networks. Previous work found that LayerCAM is a reliable visualization tool for deep speaker models. In this paper, we use LayerCAM to analyze the widely-adopted data augmentation (DA) approach, to understand how it leads to model robustness. We conduct experiments on the VoxCeleb1 dataset for speaker identification, which shows that both vanilla and activation-based (Act) DA approaches enhance robustness against interference, with Act DA being consistently superior. Visualization with LayerCAM suggests DA helps models learn to delete temporal-frequency (TF) bins that are corrupted by interference. The ‘learn to delete’ behavior explained why DA models are more robust than clean models, and why the Act DA is superior over the vanilla DA when the interference is nontarget speech. 
However, LayerCAM still cannot clearly explain the superiority of Act DA in other situations, suggesting further research.

\end{abstract}
\noindent\textbf{Index Terms}: speaker recognition, visualization, data augmentation

\section{Introduction}

Despite the great advance recently obtained in the field of speaker recognition~\cite{dehak2010front,ehsan14,li2017deep,snyder2018x,cai2018exploring},
ambient noise and other interference severely degrade the performance of speaker recognition systems~\cite{rao2014robust,zheng2017robustness}.
To improve robustness, modern deep speaker models, e.g., the x-vector \cite{snyder2018x}, extensively rely on data augmentation (DA).
Popular DA methods include gain augmentation, noise mixing, reverberation simulation, time stretch, and SpecAug~\cite{park2019specaugment,wang2020investigation}.
These DA methods have been widely used by nearly all speaker recognition systems~\cite{novotny2018use,zhao2021speakin,li2022cnsrc,lin2022robust} and implemented in most known toolkits,
e.g, Kaldi~\cite{povey2011kaldi} and Speechbrain~\cite{ravanelli2021speechbrain}.

The core idea of DA is to increase the size and diversity of the training set by generating new samples from the existing training data.
With the newly generated data, the model is supposed to learn speaker patterns in more complex conditions, hence leading to increased
robustness. Although the main principle of DA is widely adopted, \emph{we still have not a clear understanding of how DA models behave differently from clean models, and
in which way the DA models gain robustness against interference.}
A lack of such understanding prevents us from interpreting the output of DA models (e.g., confidence estimation),
designing smart verification/identification protocols (e.g., repetition policy), and constructing more efficient DA processes.
Unfortunately, deciphering deep speaker models turns out to be highly difficult, as neural networks are notorious for their black-box nature.

Recently, various visualization methods have been developed to understand the internal mechanisms of deep models,
and remarkable success has been achieved particularly in the computer vision field~\cite{simonyan2014very,ribeiro2016should,zhou2016learning}.
For speaker recognition, there are also some attempts to use such visualization tools to explain deep speaker models.
For example, Zhou et al.~\cite{zhou2021resnext} used Grad-CAM~\cite{selvaraju2017grad} to compare ResNet and Res2Net.
They found that the saliency map produced by Grad-CAM is more stable with Res2Net compared with ResNet,
thus explaining the advantage of Res2Net.
In another work, Himawan et al.~\cite{himawan2019voice} used Grad-CAM to analyze the difference between
genuine and spoof speech from the perspective of deep CNN anti-spoof models. They found that the CNN model
identifies spoof speech by looking into high-frequency components.
A potential problem of these studies is that the visualization tools are used `off-the-shelf',
without inquiring if the explanation generated by these tools is correct.
Recent research~\cite{li2022reliable} showed a surprising discovery: blindly using the visualization tools may produce a misleading
explanation for deep speaker models. The authors studied various algorithms in the CAM family, and found that only LayerCAM~\cite{jiang2021layercam} can identify
the important temporal-frequency (TF) bins and offer a reliable explanation.

In this paper, we use LayerCAM as the probing tool to understand the contribution of data augmentation in training deep speaker models.
To simplify the study, we focus on a particular and mostly used DA approach --- interference mixing,
and test only known interference. We use LayerCAM to study two DA algorithms:
vanilla DA and activation-based DA (Act DA). The latter
involves an additional loss that enforces augmented speech close to the original clean speech in the embedding space.

A speaker identification (SID) experiment was designed to test and analyze the DA methods, using the VoxCeleb1 dataset.
The results show that DA contributes by letting the models `learn to delete', i.e., learn to detect temporal-frequency (TF)
bins belonging to or corrupted by interference. This `learn to delete' hypothesis provides a possible explanation for the
DA approach and sheds more light on how deep speaker models gain their decisions from speech signals.

\section{Revisit LayerCAM}

LayerCAM~\cite{li2022reliable,jiang2021layercam} is a vital tool for visualizing CNN models. It constructs a \emph{saliency map} of the same size as the original
input (e.g., a picture or a Mel spectrogram). This saliency map shows the important regions when a CNN model tries to identify a particular class.

Let $f$ denote the speaker classifier instantiated by a CNN, and $\theta$ represents its parameters.
For a given input $x$ from class $c$, the prediction score (posterior probability) for the target class can be computed by a forward pass:
\vspace{-1.0mm}
\begin{equation}
\label{eq:pred}
y^c = f_c(x; \theta).
\end{equation}
Secondly, the weight for $k$-th activation map $A^k$ for class $c$ at location ($i,j$) is defined as the gradient at that location:
\begin{equation}
\label{eq:layer-w}
w_{ij}^{kc} = \text{ReLU } (\frac{\partial y^{c}}{\partial A_{i j}^{k}}).
\end{equation}
\noindent Finally, the saliency map is produced as follows:
\begin{equation}
  \label{eq:layer}
  S_{ij}^c = \text{ReLU } ( \sum_k w_{ij}^{kc} \cdot {A_{i j}^{k}} ).
\end{equation}
\noindent We normalize $S_{ij}^c$ to the range [0,1] following the same procedure recommended in~\cite{li2022reliable}.
Moreover, $S_{ij}^c$ is computed for each CNN layer, and we fuse $S_{ij}^c$ of four layers (ResNetBlock1-ResNetBlock4) by element-wise average manner,
as suggested in~\cite{li2022reliable}.

\section{Data augmentation}

In this paper, we use LayerCAM as a visualization tool to analyze two DA variants: the vanilla DA and the Act DA which involves an additional loss in the embedding space.

\subsection{Vanilla DA}

For the Vanilla DA, we randomly sample interference signals from MUSAN~\cite{snyder2015musan} dataset and mix it with the target speech.
Let $x$ denote clean speech, $\tilde{x}$ is an augmented version derived from $x$ by:
\begin{equation}
\tilde{x} = x + \alpha n   \ \ \ \ \ \ \alpha \sim \text{Uniform} (0.1, 2.0)
\end{equation}
\noindent where $n$ is an interference signal sampled from the MUSAN dataset. The loss function is written by:
\begin{equation}
L_{\text{Vanilla-DA}} = L(x) + L(\tilde{x})
\end{equation}
\noindent where $L$ is the objective function used for training the deep speaker model, which is the cross-entropy loss in our study.
There are three types of interference in MUSAN: noise, speech, and music. 
As mentioned, we train and test DA models for the three interference separately,
with the sake of simplifying the analysis.

\subsection{Act DA}

We design a variant of the vanilla DA named Act DA. It introduces an extra constraint that enforces the clean and augmented speech
close to each other in the embedding space. Let $e(x)$ denote the embedding of $x$, the loss function of the Act DA is as follows:

\begin{equation}
\label{equ:activation-based}
L_{\text{Act-DA}} = L(x) + L(\tilde{x})+ ||e(x) - e(\tilde{x})||^2
\end{equation}

\section{Experiment}

In this section, we first validate the robustness of the two DA methods with a speaker identification task,
and secondly apply LayerCAM to explain how DA improves the robustness of deep speaker models.

\subsection{Power of data augmentation}
\label{section:exp1}

In this section, we design a speaker identification (SID) task to demonstrate the power of data augmentation.
The structure of the model is ResNet34L with squeeze-and-excitation (SE) layers~\cite{hu2018squeeze}.
The dataset used to conduct the experiments is Voxceleb1~\cite{nagrani2017voxceleb}, and we use the
standard development set to train the model and the evaluation set for SID to test the performance.
There are 1,251 speakers in total.
The training is conducted following the voxceleb/v2 recipe of the Sunine toolkit\footnote{https://gitlab.com/csltstu/sunine}.
Note that the speakers in the test set also appear in the training set, so the outputs of the ResNet34L network
are used directly to identify the target speaker.

The MUSAN database is used to sample interference signals, including three types: noise, speech, and music.
As mentioned, we train DA models with each type of interference,
and test the models on speech with the same interference.
This allows us to focus on how a DA model learns to deal with interference it sees during training. 
More complicated cross-condition generalization will be left for future work.

Table~\ref{tab:sid} shows the Top-1/5/10 accuracy of the models trained/tested with each type of interference.
It can be seen that the clean model can obtain pretty good performance on clean speech, however, it
suffers significant performance degradation on noisy speech no matter which type of interference. The DA models
demonstrate remarkable performance improvement on noisy speech without much impact on clean speech.
Interestingly, the Act DA model shows a more significant improvement than the vanilla DA model, and the advantage is more
evident in the scenario with speech interference.

\begin{table}[!htpb]
\centering
\caption{Accuracy(\%) of models tested under different interference conditions, with/without data augmentation. Note that
for noise/speech/music conditions, DA models are trained with matched interference; for the clean condition, the results with DA models are the mean of the three DA models trained with different interference.}
\vspace{-1mm}
\label{tab:sid}
\scalebox{0.83}{
\begin{tabular}{llccc}
\toprule
\multicolumn{2}{l}{Conditions}    & Base  & Vanilla DA & Act DA \\ \cmidrule(r){1-2} \cmidrule(r){3-3} \cmidrule(r){4-4}  \cmidrule(r){5-5}
\multirow{3}{*}{Clean}  & Top-1  & 96.64 & 95.68      & 95.35  \\ 
                        & Top-5  & 98.80 & 98.63      & 98.49  \\
                        & Top-10 & 99.21 & 99.09      & 98.97  \\ \cmidrule(r){1-2} \cmidrule(r){3-3} \cmidrule(r){4-4}  \cmidrule(r){5-5}
\multirow{3}{*}{Noise}  & Top-1  & 47.30 & 73.40      & 75.02  \\ 
                        & Top-5  & 57.30 & 82.25      & 83.74  \\
                        & Top-10 & 61.23 & 85.09      & 86.75  \\ \cmidrule(r){1-2} \cmidrule(r){3-3} \cmidrule(r){4-4}  \cmidrule(r){5-5}
\multirow{3}{*}{Speech} & Top-1  & 42.31 & 75.53      & 81.08  \\
                        & Top-5  & 53.96 & 86.23      & 91.10  \\
                        & Top-10 & 58.24 & 89.44      & 93.63  \\  \cmidrule(r){1-2} \cmidrule(r){3-3} \cmidrule(r){4-4}  \cmidrule(r){5-5}
\multirow{3}{*}{Music}  & Top-1  & 26.47 & 49.74      & 53.23  \\
                        & Top-5  & 35.05 & 61.37      & 64.68  \\
                        & Top-10 & 38.71 & 65.47      & 68.25  \\ 
\bottomrule
\end{tabular}}
\end{table}

The above results confirm that DA training can lead to a significant improvement in model robustness, and the Act DA training
leads to further gains. However, how the improvement is obtained by DA and how Act DA wins vanilla DA is still vague.
We will apply LayerCAM to get a deeper understanding.

\subsection{Concatenated interference}
\label{sec:concate}

\begin{figure}[!htbp]
\includegraphics[width=0.96\linewidth]{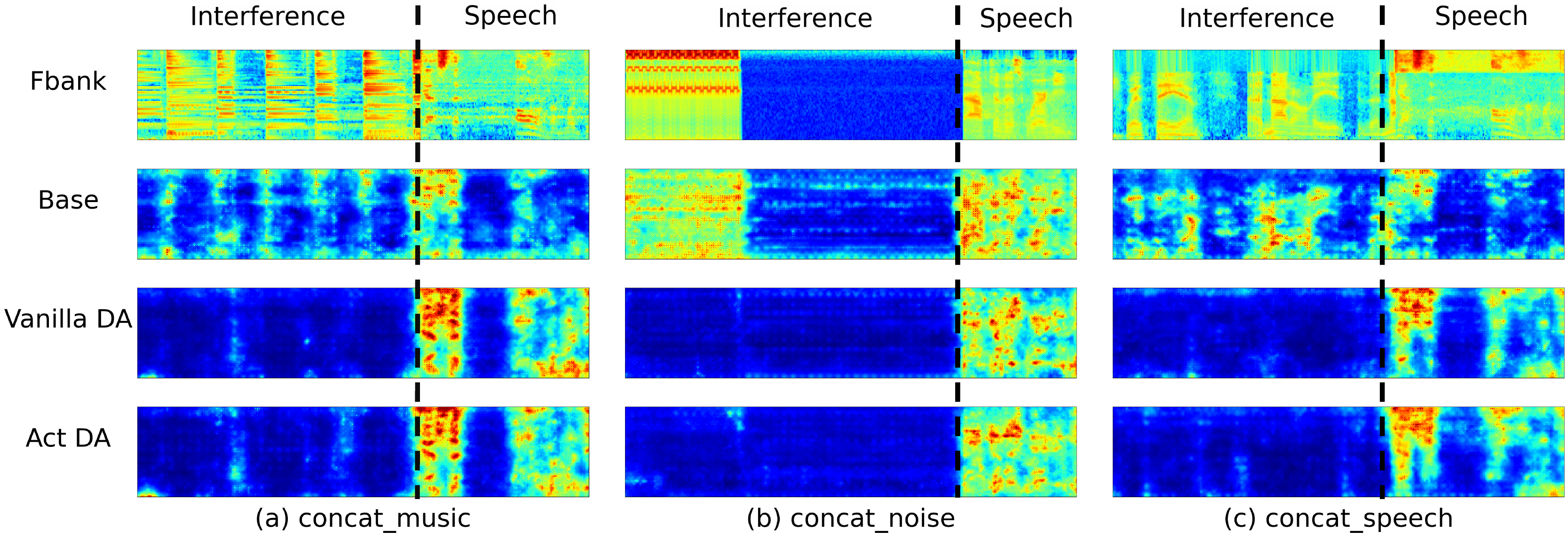}
\centering
\vspace{-1mm}
\caption{Saliency maps of three utterances sampled the test set and concatenated with different types of interference.
The first row shows the Fbank features, and the rest rows show the saliency maps produced by LayerCAM with the clean model (2nd row), the
vanilla DA model (3rd row) and the Act DA model (4th row).
The saliency values have been scaled to the range [0,1], and high values represent important regions and are shown by warm color.}
\label{fig:concat}
\end{figure}

We start from a simple scenario where the target speech and the interference are concatenated.
The saliency maps found by LayerCAM are shown in Figure~\ref{fig:concat}, where each column is a speech that is sampled from the
test set and concatenated by a particular interference. It can be seen that both the clean model and the two DA models
can identify the interference segment, while the DA models (3rd and 4th rows) work much better than the clean model (2nd row).

To quantify the performance of different models in identifying interference regions, we compute
the speech preservation ratio (SPR) and interference preservation ratio (IPR) with the VoxCeleb1 test set.
Specifically, we sum the saliency values across the frequency axis to get the saliency value for each frame,
and if the value is above 15 (empirical set) the frame
is regarded to belong to the target speech, otherwise, it belongs to the interference.
SPR and IPR are defined as the retention ratio of the target speech and the interference signal after the above dichotomy, respectively.

\begin{table}[!ht]
\caption{Quality of saliency maps produced by LayerCAM with different models. SPR: speech preservation ratio; IPR: interference preservation ratio.}
\vspace{-1mm}
\label{tab:concat}
\centering
\scalebox{0.83}{
\begin{tabular}{llccc}
\toprule
 \multicolumn{2}{l}{Conditions}   & Base  & Vanilla DA   & Act DA        \\ \cmidrule(r){1-2} \cmidrule(r){3-3} \cmidrule(r){4-4} \cmidrule(r){5-5}
\multirow{3}{*}{SPR($\uparrow$)} & Noise  & 95.8\% & 93.4\%         & 94.2\%          \\ 
                     & Speech & 92.8\% & 94.9\%         & 94.2\%          \\
                     & Music  & 95.6\% & 92.5\%         & 93.0\%          \\ \cmidrule(r){1-2} \cmidrule(r){3-3} \cmidrule(r){4-4} \cmidrule(r){5-5}
\multirow{3}{*}{IPR($\downarrow$)} & Noise  & 42.3\% & \textbf{5.5\%} & 9.5\%           \\
                     & Speech & 81.9\% & 47.7\%         & \textbf{25.7\%} \\
                     & Music  & 43.7\% & \textbf{5.2\%} & 8.3\%           \\ 
\bottomrule
\end{tabular}}
\end{table}

The results are shown in Table~\ref{tab:concat}. First of all, the SPR values of all the models are not
significantly different, suggesting that the speech segments are all well identified by all the models.
However, the IPR values show much difference: the DA models identify the interference much more successfully than
the clean model, indicating that by DA training, the models have learned what is useless. We, therefore, conjecture that
the main role that DA plays is to learn how to delete interference, instead of identifying speech segments.
This trend is most clear in the test with the speech interference, where the DA models are much stronger than the clean model in
identifying and removing the interference, and the Act DA model is
substantially stronger than the vanilla DA model.
Considering the superior performance of Act DA in the SID experiment, we hypothesize that the `learn to delete’
feature contributes the most to the success of DA training.

\subsection{Overlapped interference}

In this section, we consider a more complicated scenario where the target speech and the interference are overlapped. 
We start by visualizing the saliency maps and then conduct quantitative analysis.

\subsubsection{Saliency map}

\begin{figure}[!htbp]
\centering
\includegraphics[width=0.96\linewidth]{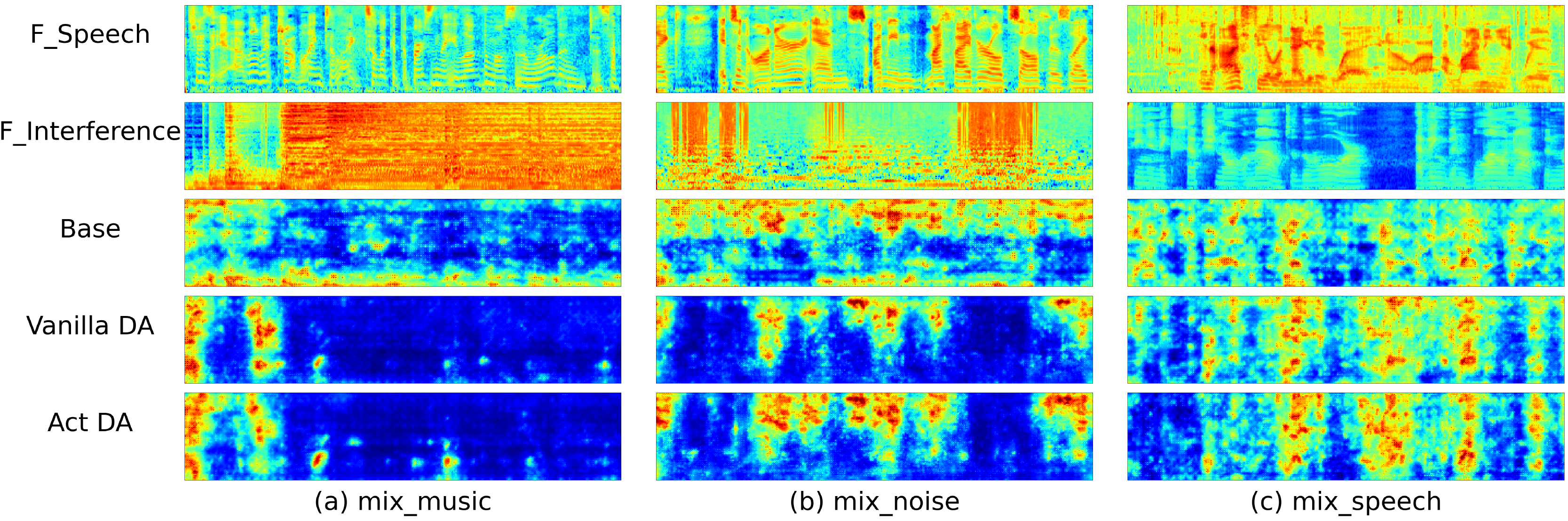}
\vspace{-1mm}
\caption{Saliency maps of speech overlapped with three types of interference (music, noise, speech). The first and second rows are Fbank features of
the target speech and interference, and the
rest rows are the saliency maps produced by LayerCAM for the clean model (3rd row), vanilla DA model (4th row), and Act DA model (5th row). }
\label{fig:mixed}
\vspace{-2mm}
\end{figure}

The saliency maps of three samples overlapped with different types of interference are shown in Figure~\ref{fig:mixed}.
To make the presentation clear, the Fbank features of the target speech and the interference are shown separately in the
first and the second row.

We draw some similar conclusions as in the scenario with concatenated interference. For example, the DA models are more powerful in identifying
the interference region than the clean model, and the Act DA model works better than the vanilla DA model under the condition
with speech interference, as it identifies more TF bins with strong interference. This is consistent with the
low IPR value that the Act DA model obtains in Table~\ref{tab:concat}. Perhaps the most important observation is that with the two DA models,
high saliency values are assigned to the TF regions where the interference is weak, rather than the regions where the target speech is strong.
This double confirms the `learn to delete' hypothesis we proposed in Section~\ref{sec:concate}.

\begin{figure*}[!htbp]
  \centering
  \includegraphics[width=0.93\linewidth]{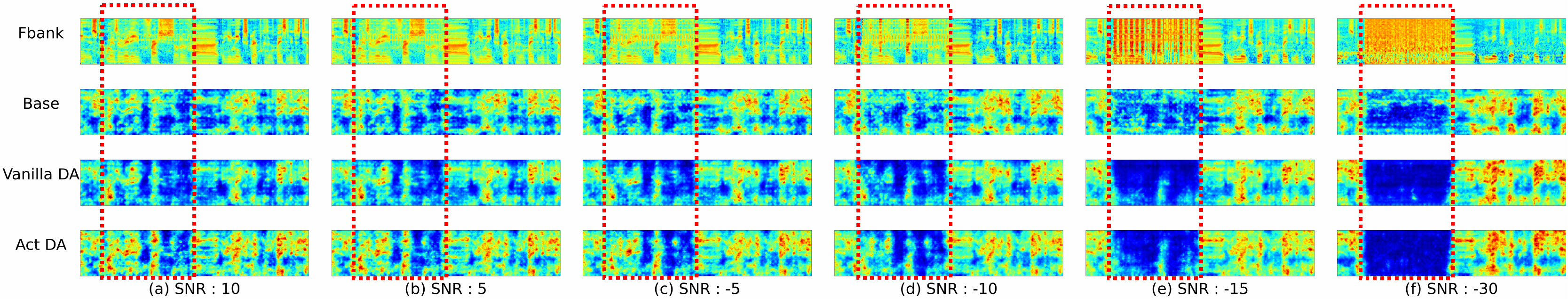}
  \vspace{-1mm}
  \caption{Fbank features (1st row) and saliency maps from the clean model (2nd row) and the two DA models (3rd and 4th rows), when a
  phone ring is mixed with the target speech in the areas shown by the red boxes. The intensity of the noise is increased gradually from left to right.}
  \label{fig:exp4}
  \vspace{-1mm}
\end{figure*}

\subsubsection{Quantity analysis}

Since the target speech and interference are overlapped, the SPR/IPR metrics are not suitable anymore for the quantitative analysis.
To solve the problem, we borrow some metrics from speech enhancement research.
Specifically, the saliency values are treated as soft masks to `denoise' the interfered speech. This can be simply conducted by
multiplying the saliency values to the Fbank features of the noisy data and then converting the denoised features
to waveforms. The conversion is based on inverse FFT, with the phase spectrogram of the clean speech.
For a fair comparison, the noisy speech also undergoes the same re-synthesizing process. We
then measure the quality of the denoised speech by PESQ, STOI, and SNR, three popular metrics in speech enhancement, which we assume
reflect the quality of the saliency maps.

\begin{table}[!htbp]
\caption{Quality of the denoised speech when applying the saliency values as the denoising masks.}
\vspace{-1mm}
\label{table:speech_qua}
\centering
\scalebox{0.8}{
\begin{tabular}{llcccc}
\toprule
 \multirow{2}{*}{Conditions} &  & \multirow{2}{*}{Noisy}  & \multicolumn{3}{c}{Saliency Mask}    \\  \cmidrule(r){4-6} 
 &  &   & Base   & Vanilla DA    & Act DA        \\ \cmidrule(r){1-2} \cmidrule(r){3-3} \cmidrule(r){4-4} \cmidrule(r){5-5} \cmidrule(r){6-5}
\multirow{3}{*}{Noise}        & PESQ ($\uparrow$)                           & 1.430  & 1.442  & \textbf{1.473}  & 1.472  \\
                              & STOI ($\uparrow$)                           & 0.688  & 0.689  & \textbf{0.707}  & \textbf{0.707}  \\
                              & SNR ($\uparrow$)                            & -5.358 & -4.850 & -4.602          & \textbf{-4.548} \\ \cmidrule(r){1-2} \cmidrule(r){3-3} \cmidrule(r){4-4} \cmidrule(r){5-5} \cmidrule(r){6-5}
\multirow{3}{*}{Speech}       & PESQ ($\uparrow$)                          & 1.377  & 1.372  & 1.387  & \textbf{1.392}  \\
                              & STOI ($\uparrow$)                          & 0.616  & 0.607  & 0.628           & \textbf{0.640}  \\
                              & SNR ($\uparrow$)                           & -5.648 & -5.313 & -5.129          & \textbf{-4.810} \\ \cmidrule(r){1-2} \cmidrule(r){3-3} \cmidrule(r){4-4} \cmidrule(r){5-5} \cmidrule(r){6-5}
\multirow{3}{*}{Music}        & PESQ ($\uparrow$)                          & 1.247  & 1.249  & \textbf{1.254}  & 1.252           \\
                              & STOI ($\uparrow$)                          & 0.529  & 0.529  & \textbf{0.557}  & 0.556           \\
                              & SNR ($\uparrow$)                           & -8.277 & -7.673 & \textbf{-7.027} & -7.141          \\ 
\bottomrule
\end{tabular}}
\end{table}

The results are presented in Table~\ref{table:speech_qua}.
It can be seen that denoising with saliency maps from the clean model (2nd column) indeed leads to marginal but consistent improvement in speech quality, especially
in terms of SNR. This indicates that the clean model can detect the interference TF bins to some extent.
In comparison, denoising with the saliency maps from the augmented models leads to more significant improvement, indicating that DA models are more powerful in
detecting and removing interference. By comparing the two DA models, the Act DA model shows an advantage when the interference is speech, which is consistent with
the results in Table~\ref{tab:concat} and Figure~\ref{fig:mixed}.

\subsection{Corrupted speech}

In the last experiment, we show how clean and DA models treat the speech segments corrupted by
interference: will they be still used anyway or simply thrown away?
Figure~\ref{fig:exp4} presents a show-case study, where we mix a noise speech (a phone ring) with a target speech and increase the intensity of the
noise gradually. The Fbank features of the noisy speech and the saliency maps from different models are shown.
Note that the red boxes indicate the speech segment corrupted by the phone ring.

It can be seen that when the noise is weak (SNR = 10, 5), most of the TF bins of the corrupted speech remain, 
even though some information has been lost due to corruption.
When the noise becomes stronger (SNR = -5, -10, -15), the noise has a more impact, especially on the high-frequency region
where the energy of human speech is weak~\cite{monson2012analysis}. In this scenario, the saliency values of the high-frequency TF bins
are reduced, though the low-frequency bins are still retained. When the noise is very strong (SNR = -30),
the saliency values of the whole noisy segment are reduced to zero.
The above trend is similar with both the clean model and the two DA models, though it is much more clear
with the DA models. This suggests that DA models not only learn to delete interference but also learn to
delete speech regions if they are severely corrupted.

A deletion test~\cite{li2022reliable,petsiuk2018rise} is conducted to verify this behavior quantitatively. In this test,
we firstly compute the saliency maps of \emph{noisy} data, and progressively mask the TF bins of the \emph{clean} speech whose saliency values in
TF bins of the \emph{noisy} speech are small.
Finally, we test the SID accuracy with the masked clean speech using the \emph{clean} model.
If a model tends to delete corrupted speech segments, the performance tested on the masked clean speech with the clean model will drop significantly, as
the deleted segments are informative for the clean model.

\begin{figure}[!h]
  \centering
  \includegraphics[width=0.98\linewidth]{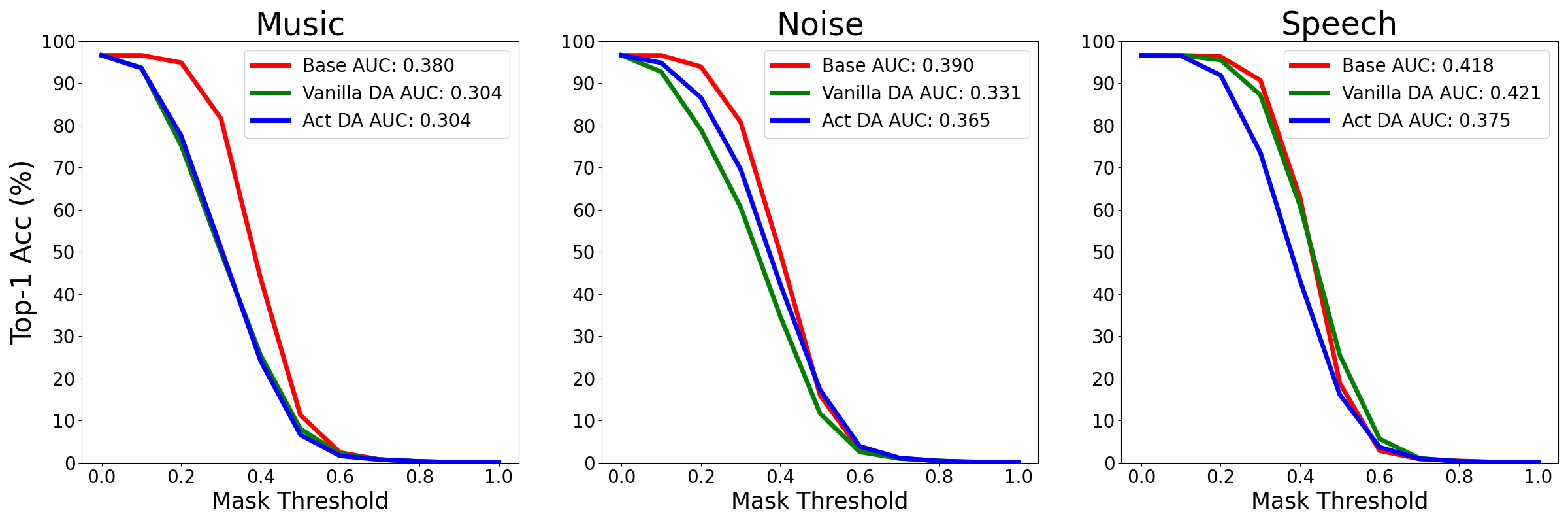}
  \vspace{-0.5mm}
  \caption{Deletion test to verify that DA models tend to delete more corrupted speech compared to clean models. Clean speech is masked according to the
  saliency map of noisy speech and then tested with the clean model. Gradually increasing the threshold on the saliency values to be masked gives the curves shown in the three plots. AUC: Area Under the Curve.}
  \label{fig:auc}
  \vspace{-0.5mm}
\end{figure}

The results are shown in Figure~\ref{fig:auc}, where each plot shows the SID accuracy when the threshold on the saliency values for the TF bin masking is increased.
It can be seen that the DA models exhibit a more significant performance drop, indicating that they delete some speech TF bins that would be useful for the clean model.
An interesting observation is that the Act DA model tends to be more aggressive in masking corrupted speech in the condition with speech interference (right plot).
This is also consistent with the observation in Figure~\ref{fig:exp4} and explained why Act DA performs significantly better than vanilla DA in that condition as shown in Table~\ref{tab:sid}.

\section{Conclusions}

This paper aims to understand the role that data augmentation plays in speaker recognition, by using LayerCAM as a visualization tool.
By speaker identification experiments conducted with the VoxCeleb1 dataset, we found that data augmentation functions by
detecting interference regions as well as speech regions corrupted by strong interference, and removing their impact by assigning
low saliency values in these regions. We call this the `learn to delete' hypothesis. This hypothesis explains how data augmentation
improves model robustness, and explains why an activation-based DA performs better than the vanilla DA when the interference signal
is speech.

We admit that the `explanation' is still superficial and far from a full understanding of the working mechanism of deep speaker models.
For instance, how the model discovers speaker-related patterns and how DA impacts such discovery thus leading to robustness. Future work
will focus on a more zoom-in understanding of the model and more complex robustness
that various DA methods and other advanced techniques have contributed.



\bibliographystyle{IEEEtran}
\bibliography{mybib}

\begin{thebibliography}{10}
\providecommand{\url}[1]{#1}
\csname url@samestyle\endcsname
\providecommand{\newblock}{\relax}
\providecommand{\bibinfo}[2]{#2}
\providecommand{\BIBentrySTDinterwordspacing}{\spaceskip=0pt\relax}
\providecommand{\BIBentryALTinterwordstretchfactor}{4}
\providecommand{\BIBentryALTinterwordspacing}{\spaceskip=\fontdimen2\font plus
\BIBentryALTinterwordstretchfactor\fontdimen3\font minus
  \fontdimen4\font\relax}
\providecommand{\BIBforeignlanguage}[2]{{%
\expandafter\ifx\csname l@#1\endcsname\relax
\typeout{** WARNING: IEEEtran.bst: No hyphenation pattern has been}%
\typeout{** loaded for the language `#1'. Using the pattern for}%
\typeout{** the default language instead.}%
\else
\language=\csname l@#1\endcsname
\fi
#2}}
\providecommand{\BIBdecl}{\relax}
\BIBdecl

\bibitem{dehak2010front}
N.~Dehak, P.~J. Kenny, R.~Dehak, P.~Dumouchel, and P.~Ouellet, ``Front-end
  factor analysis for speaker verification,'' \emph{IEEE Transactions on Audio,
  Speech, and Language Processing}, vol.~19, no.~4, pp. 788--798, 2010.

\bibitem{ehsan14}
E.~Variani, X.~Lei, E.~McDermott, I.~L. Moreno, and J.~Gonzalez-Dominguez,
  ``Deep neural networks for small footprint text-dependent speaker
  verification,'' in \emph{IEEE International Conference on Acoustics, Speech
  and Signal Processing (ICASSP)}.\hskip 1em plus 0.5em minus 0.4em\relax IEEE,
  2014, pp. 4052--4056.

\bibitem{li2017deep}
L.~Li, Y.~Chen, Y.~Shi, Z.~Tang, and D.~Wang, ``Deep speaker feature learning
  for text-independent speaker verification,'' in \emph{INTERSPEECH}, 2017, pp.
  1542--1546.

\bibitem{snyder2018x}
D.~Snyder, D.~Garcia-Romero, G.~Sell, D.~Povey, and S.~Khudanpur, ``X-vectors:
  Robust {DNN} embeddings for speaker recognition,'' in \emph{IEEE
  International Conference on Acoustics, Speech and Signal Processing
  (ICASSP)}.\hskip 1em plus 0.5em minus 0.4em\relax IEEE, 2018, pp. 5329--5333.

\bibitem{cai2018exploring}
W.~Cai, J.~Chen, and M.~Li, ``Exploring the encoding layer and loss function in
  end-to-end speaker and language recognition system,'' \emph{arXiv preprint
  arXiv:1804.05160}, 2018.

\bibitem{rao2014robust}
K.~S. Rao and S.~Sarkar, \emph{Robust speaker recognition in noisy
  environments}.\hskip 1em plus 0.5em minus 0.4em\relax Springer, 2014.

\bibitem{zheng2017robustness}
T.~F. Zheng and L.~Li, \emph{Robustness-related issues in speaker
  recognition}.\hskip 1em plus 0.5em minus 0.4em\relax Springer, 2017, vol.~2.

\bibitem{park2019specaugment}
D.~S. Park, W.~Chan, Y.~Zhang, C.-C. Chiu, B.~Zoph, E.~D. Cubuk, and Q.~V. Le,
  ``Specaugment: A simple data augmentation method for automatic speech
  recognition,'' in \emph{INTERSPEECH}, 2019, pp. 2613--2617.

\bibitem{wang2020investigation}
S.~Wang, J.~Rohdin, O.~Plchot, L.~Burget, K.~Yu, and J.~{\v{C}}ernock{\`y},
  ``Investigation of specaugment for deep speaker embedding learning,'' in
  \emph{IEEE International Conference on Acoustics, Speech and Signal
  Processing (ICASSP)}.\hskip 1em plus 0.5em minus 0.4em\relax IEEE, 2020, pp.
  7139--7143.

\bibitem{novotny2018use}
O.~Novotn{\`y}, O.~Plchot, P.~Matejka, L.~Mosner, and O.~Glembek, ``On the use
  of x-vectors for robust speaker recognition.'' in \emph{Odyssey}, 2018, pp.
  168--175.

\bibitem{zhao2021speakin}
M.~Zhao, Y.~Ma, M.~Liu, and M.~Xu, ``The speakin system for voxceleb speaker
  recognition challange 2021,'' \emph{arXiv preprint arXiv:2109.01989}, 2021.

\bibitem{li2022cnsrc}
\BIBentryALTinterwordspacing
L.~Li, ``Technical overview for {CNSRC} 2022,'' Tsinghua University, 2022.
  [Online]. Available: \url{http://cnceleb.org/workshop}
\BIBentrySTDinterwordspacing

\bibitem{lin2022robust}
W.~Lin and M.-W. Mak, ``Robust speaker verification using population-based data
  augmentation,'' in \emph{IEEE International Conference on Acoustics, Speech
  and Signal Processing (ICASSP)}.\hskip 1em plus 0.5em minus 0.4em\relax IEEE,
  2022, pp. 7642--7646.

\bibitem{povey2011kaldi}
D.~Povey, A.~Ghoshal, G.~Boulianne, L.~Burget, O.~Glembek, N.~Goel,
  M.~Hannemann, P.~Motlicek, Y.~Qian, P.~Schwarz \emph{et~al.}, ``The {K}aldi
  speech recognition toolkit,'' in \emph{IEEE 2011 workshop on automatic speech
  recognition and understanding}, no. CONF.\hskip 1em plus 0.5em minus
  0.4em\relax IEEE Signal Processing Society, 2011.

\bibitem{ravanelli2021speechbrain}
M.~Ravanelli, T.~Parcollet, P.~Plantinga, A.~Rouhe, S.~Cornell, L.~Lugosch,
  C.~Subakan, N.~Dawalatabad, A.~Heba, J.~Zhong \emph{et~al.}, ``{SpeechBrain}:
  A general-purpose speech toolkit,'' \emph{arXiv preprint arXiv:2106.04624},
  2021.

\bibitem{simonyan2014very}
K.~Simonyan and A.~Zisserman, ``Very deep convolutional networks for
  large-scale image recognition,'' \emph{arXiv preprint arXiv:1409.1556}, 2014.

\bibitem{ribeiro2016should}
M.~T. Ribeiro, S.~Singh, and C.~Guestrin, ````{W}hy should {I} trust you?''
  explaining the predictions of any classifier,'' in \emph{Proceedings of the
  22nd ACM SIGKDD international conference on knowledge discovery and data
  mining}, 2016, pp. 1135--1144.

\bibitem{zhou2016learning}
B.~Zhou, A.~Khosla, A.~Lapedriza, A.~Oliva, and A.~Torralba, ``Learning deep
  features for discriminative localization,'' in \emph{IEEE conference on
  computer vision and pattern recognition}, 2016, pp. 2921--2929.

\bibitem{zhou2021resnext}
T.~Zhou, Y.~Zhao, and J.~Wu, ``Resnext and res2net structures for speaker
  verification,'' in \emph{IEEE Spoken Language Technology Workshop
  (SLT)}.\hskip 1em plus 0.5em minus 0.4em\relax IEEE, 2021, pp. 301--307.

\bibitem{selvaraju2017grad}
R.~R. Selvaraju, M.~Cogswell, A.~Das, R.~Vedantam, D.~Parikh, and D.~Batra,
  ``Grad-{CAM}: Visual explanations from deep networks via gradient-based
  localization,'' in \emph{Proceedings of the IEEE international conference on
  computer vision}, 2017, pp. 618--626.

\bibitem{himawan2019voice}
I.~Himawan, S.~Madikeri, P.~Motlicek, M.~Cernak, S.~Sridharan, and C.~Fookes,
  ``Voice presentation attack detection using convolutional neural networks,''
  \emph{Handbook of Biometric Anti-Spoofing: Presentation Attack Detection},
  pp. 391--415, 2019.

\bibitem{li2022reliable}
P.~Li, L.~Li, A.~Hamdulla, and D.~Wang, ``Reliable visualization for deep
  speaker recognition,'' in \emph{INTERSPEECH}, 2022, pp. 331--335.

\bibitem{jiang2021layercam}
P.-T. Jiang, C.-B. Zhang, Q.~Hou, M.-M. Cheng, and Y.~Wei, ``Layer{CAM}:
  Exploring hierarchical class activation maps for localization,'' \emph{IEEE
  Transactions on Image Processing}, vol.~30, pp. 5875--5888, 2021.

\bibitem{snyder2015musan}
D.~Snyder, G.~Chen, and D.~Povey, ``Musan: A music, speech, and noise corpus,''
  \emph{arXiv preprint arXiv:1510.08484}, 2015.

\bibitem{hu2018squeeze}
J.~Hu, L.~Shen, and G.~Sun, ``Squeeze-and-excitation networks,'' in
  \emph{Proceedings of the IEEE conference on computer vision and pattern
  recognition}, 2018, pp. 7132--7141.

\bibitem{nagrani2017voxceleb}
A.~Nagrani, J.~S. Chung, and A.~Zisserman, ``Vox{C}eleb: A large-scale speaker
  identification dataset,'' in \emph{INTERSPEECH}, 2017, pp. 2616--2620.

\bibitem{monson2012analysis}
B.~B. Monson, B.~Story, and A.~Lotto, ``Analysis of high-frequency energy in
  singing and speech,'' \emph{The Journal of the Acoustical Society of
  America}, vol. 131, no.~4, pp. 3378--3378, 2012.

\bibitem{petsiuk2018rise}
V.~Petsiuk, A.~Das, and K.~Saenko, ``Rise: Randomized input sampling for
  explanation of black-box models,'' \emph{arXiv preprint arXiv:1806.07421},
  2018.

\end{thebibliography}

\end{document}